%% file: main.tex
\documentclass[sigconf]{acmart}
\AtBeginDocument{%
  }

\usepackage{graphicx}   %
\usepackage{subcaption}
\usepackage{amsmath}
\usepackage{xspace}
\usepackage{makecell}
\usepackage{stfloats}
\usepackage{xcolor}
\usepackage[table]{xcolor}

\newif\ifextended
\extendedfalse  %

\newcommand{\high}{\cellcolor{green!20}High}
\newcommand{\low}{\cellcolor{red!20}Low}
\newcommand{\partialv}{\cellcolor{orange!25}Partial}
\newcommand{\nonev}{\cellcolor{gray!20}None}

\input{sections/notations} %

\newif\ifshow
\showfalse   %

\newcommand{\showif}[1]{\ifshow #1 \fi}

\begin{document}

\copyrightyear{2026}
\acmYear{2026}
\setcopyright{cc}
\setcctype{by-nc-nd}
\acmConference[SeQureDB '26]{Workshop on Secure and Private Data Management}{May 31-June 05, 2026}{Bengaluru, India}
\acmBooktitle{Workshop on Secure and Private Data Management (SeQureDB '26), May 31-June 05, 2026, Bengaluru, India}
\acmDOI{10.1145/3807894.3810276}
\acmISBN{979-8-4007-2219-6/2026/05}

\title{Policy-aware Vector Search: A Vision for Fine Grained Access Control in Vector Databases}

\author{Lakshmi Sahithi Yalamarthi}
\email{yalam2@pdx.edu}
\affiliation{%
  \institution{Portland State University}
  \city{Portland}
  \state{Oregon}
  \country{USA}
}

\author{Primal Pappachan}
\email{primal@pdx.edu}
\affiliation{%
  \institution{Portland State University}
  \city{Portland}
  \state{Oregon}
  \country{USA}
}

\input{sections/abstract}



\maketitle

\input{sections/introduction}

\input{sections/models}

\input{sections/approach}

\input{sections/experiments}

\input{sections/discussion}

\input{sections/conclusions}

\bibliographystyle{ACM-Reference-Format}
\bibliography{references}

\end{document}

%% file: sections/notations.tex
\newcommand{\pgsql}{\textsc{PostgreSQL}\xspace}
\newcommand{\pgvector}{\textit{pgvector}\xspace}

\newcommand\ie{\textit{i.e.,\ }}

\newcommand{\Database}{\ensuremath{\mathcal{D}}}

\newcommand{\PolicySet}[1]{\ensuremath{\mathcal{P}}_{#1}}

\newcommand{\selectivity}[1]{\sigma{#1}}

\newcommand{\rhovar}{\ensuremath{\rho}}  %

%% file: sections/abstract.tex
\begin{abstract}
Vector databases are increasingly used in security sensitive contexts
with Retrieval Augmented Generation and organizational AI
pipelines; however, their security capabilities remain limited. Specifically,
Fine-grained Access Control (FGAC) which is required to
ensure that data access adheres to user-specific policies is not fully
supported in modern vector databases. Unlike relational databases,
vector databases combine structured and unstructured attributes to
provide semantic, approximate query results, which complicates
FGAC implementation. This creates an inherent tension between
enforcing FGAC policies correctly, achieving high ANN search recall
and maintaining low query latency. In this paper, we present a
vision for Policy-aware Vector Search by formalizing the FGAC policy
model in vector databases as well as the enforcement problem.
We compare various enforcement strategies, present preliminary
findings, and identify key open challenges for future research in
policy-aware vector search.

\end{abstract}

%% file: sections/introduction.tex
\section{Introduction}

Vector databases have become a foundational component of modern AI workloads, supporting applications such as semantic search, recommendation systems, and retrieval-augmented generation (RAG). As these systems are increasingly used in security sensitive environments, enforcing fine-grained access control (FGAC) over vector data is critical. This concern is especially acute in RAG pipelines, where retrieved documents directly influence generated responses from a Large Language Models. 
Current vector databases provide little to no support for the specification and enforcement of Fine-Grained Access Control (FGAC) policies.

Unlike traditional relational databases, where data is structured as rows and columns, vector databases store and query over high-dimensional embeddings using similarity search. As a result, traditional FGAC enforcement approaches developed for relational databases~\cite{sieve,tattletale} do not directly translate to vector databases.
Retrieving the relevant vectors from this space typically relies on approximate nearest neighbor (ANN) search across potentially millions of vectors. 
In addition to the embeddings, vector databases allow storing metadata attributes associated with each vector. These attributes enable filtering during query processing, typically implemented using pre-filtering or post-filtering strategies. Consequently, FGAC policies can be specified over these metadata attributes or directly over the vectors themselves. While existing metadata filtering strategies could be leveraged for FGAC enforcement, they are not well suited for this purpose. In particular, \emph{post-filtering} approaches, which first execute the vector search and then remove vectors that do not satisfy FGAC policies, can be highly inefficient when policies are selective, as many retrieved vectors may ultimately be discarded. The final recall can also be severely impacted if retrieved vectors do not satisfy the FGAC policies.

In contrast, \emph{pre-filtering} is based on Access Control Lists (ACLs) before executing ANN search. So, it may under-utilize the efficient ANN indexes such as Hierarchical Navigable Small World \cite{hnsw}, which are designed to operate over the entire vector space.
Prior works have explored partitioning the vector space and building independent indexes but this require additional space overhead of multiple indexes and maintenance of these indexes~\cite{honeybee}.
The absence of accurate cost and recall models in vector databases further complicates the choice between pre-filtering and post-filtering strategies. 
Recently emerging area of \emph{Hybrid strategies for filter-aware vector search} holds promise for FGAC enforcement.
These approaches typically necessitate index modifications~\cite{acorn,UNG}, are limited to metadata-based policies, and exhibit poor efficiency when policies affect only a narrow subset of the dataset.\cite{surveyFANN}

In this paper, we first discuss in detail a novel FGAC policy model for vector databases that
supports specification of FGAC policies on metadata attributes. We then discuss in detail the main challenge of enforcing FGAC policies that contains metadata filters.
We formally define the FGAC enforcement problem and identify the key considerations for supporting it in vector databases.
We evaluate four different approaches for FGAC policy-aware vector search on Pgvector (PostgreSQL enabled with Vector Search) to illustrate the latency-recall tradeoffs on a large real dataset with synthetic policies.
We discuss the additional challenges of FGAC management in vector databases such as storage, maintenance, and enforcing FGAC policies that are specified as vectors.
By discussing the trade-off between policy enforcement and efficient ANN retrieval, our work move towards practical, secure vector search in multi-tenant AI systems, where both retrieval quality and strict access guarantees are required.

%% file: sections/models.tex
\section{FGAC Policy Model for Vector DBs}
\label{ref:model}

In this section, we first describe the data, query, policy models and then we discuss the key considerations for implementing FGAC in vector databases.Throughout the paper, we illustrate our approach using a dataset of research papers, where vectors are generated from paper titles and abstracts~\cite{dataset}, and access-control policies are generated from other columns.

\subsubsection*{Data model:}
Let \( \mathbb{D} \) denote the database consisting of vector-metadata attribute pairs:
\[
\mathbb{D} = \{(v_1, m_1), (v_2, m_2), \dots, (v_i, m_i), \dots (v_n, m_n)\}
\]
where each vector \( v_i \in \ \mathbb{R}^{d} \) denotes a vector embedding of $d$ dimensions and $m_i$ = \(\{m^1_i, m^2_i, \dots, m^s_i\} \) denote the metadata (scalar) attributes. 
Each ${m^j_i}$ is in the key value format \(\{m^j_i : ``value"\} \) denoting the \textit{``value"} assigned to the metadata attribute $m^j_i$. For example, the \textit{category} attribute associated with a research paper is represented as \(\{categories :  ``CS"\} \). 

\subsubsection*{Query model:}

We represent a vector query as $\mathbb{Q} = \{q_v, q_m, k\}$, where $q_v \in \mathbb{R^d}$ is the vector query on the embeddings, $q_m$ is the metadata query involving zero, one, or more of the scalar attributes, and $k$ is the number of vectors to be returned as the result. The execution of $\mathbb{Q}$ over $\mathbb{D}$ returns the top-$k$ nearest vectors to $q_v$: 

\begin{equation}
V_k = \underset{S \subset \mathbb{D}, \, |S|=k}{\operatorname{arg\,min}} \sum_{v_i \in S} dist(q_v, v_i)
\end{equation}

For Example, a vector query could be ``Find papers that apply neural network architectures to solve optimization problems in resource-constrained environments'', while the metadata filters would be predicates such as \texttt{publication\_year} $\geq$ 2020. 

\subsubsection*{FGAC Policy model:}
In our work, FGAC policies are represented as a 3-tuple following the ABAC policy model~\cite{ABAC}.
\[
P_i = [oc, sc, act]
\]
\begin{itemize}
    \item \textbf{\( oc \) (Object Constraints)} identify the vector on which FGAC policy is to be enforced. 
    They are defined as a set of logical predicates defined on metadata attributes $m_i$ of a vector \( v_i \) and combined with a boolean operator: 
    $oc = ([m^1_i, \text{op}_1, \text{val}_1] \land [m^2_i, \text{op}_2, \text{val}_2] \land [m^3_i, \text{op}_3, \text{val}_3]) 
    \land \dots
    [m^s_i, \text{op}_s, \text{val}_s]) $\footnote{While object constraints can theoretically include disjunctions along conjunctions, the support for answering disjunctive metadata filters is limited in vector databases~\cite{Filtered-DiskANN}.} where
    $op \in \{=, \neq, \geq, >, \leq, < \}$ and $val \in dom(m^j_i)$.  
    From our running example, a possible set of MOCs are: (
        \text{categories} = \texttt{``CS''} \;$\land$\;
            \text{license} = \texttt{``CC-BY''} \;$\land$\;
            \text{report\_no} = \texttt{``HEP-PR-07-12''}).

    \item \textbf{\( sc \) (Subject Constraints)} identify the querier for which the FGAC policy applies.
    Similar to metadata-based object constraints, they are defined as a set of logical predicates on subject attributes of the querying user.
    Examples of subject attributes include purpose of the query~\cite{PBAC} or roles/group memberships of the user. These subject attributes are included with the query as querier metadata and used to identify the policies that apply to a given query. 
  
   \item \textbf{\( act \) (Action)} is the access decision ($\textit{act} \in \{\text{allow}, \text{deny}\}$) that is to be enforced on a vector $v_i$ that satisfies the vector query and applicable FGAC policies. 
\end{itemize}

%% file: sections/approach.tex
\section{Policy-aware Vector Search}

In this section, we briefly discuss the the various enforcement strategies for \emph{Policy-aware Vector Search}. We formalize the enforcement problem, present performance estimators, and discuss the pros and cons of different strategies. We also discuss the challenges with storage and maintenance of FGAC policies along with ideas for enforcing vector based object conditions.

\noindent \textbf{Enforcement Problem.} The FGAC enforcement problem is to ensure that query results returned to a user satisfy all applicable policy constraints associated with the user. 
Given a query \( \mathbb{Q} \) posed by a user (subject) \( S \), the set of policies that apply to the user, $\mathbb{P}_{S} = \{P_i | sc_i(S) =  \text{True}\}$.
The set of vectors, $V_P$, that satisfy the applicable policies $\mathbb{P}_{S}$ for user $S$ is given by:
\begin{equation}
\label{eq:policyCheck}
    V_{P} = \{(v, m) \in \mathbb{D} | \exists P_j \in P_{S}: P_j(v, m) = \text{True}\}     
\end{equation}
If the actions associated with the policies are \textbf{allow}, theoretically the result set $Q$ should be equivalent to retrieving the k nearest neighbors that comply with the policy filters for the query vector $q_v$ in $V_P$~\footnote{Policies with action \textit{deny} requires execution of $Q$ over $\mathbb{D} - V_P$}. 
\begin{equation}
\label{eq:preFilter}
V_Q = \underset{S \subset V_P, \, |S|=k}{\operatorname{arg\,min}} \sum_{v_i \in S} dist(q_v, v_i)
\end{equation}

\begin{table*}[htbp]
\centering
\caption{Comparison of different meta-data filtering strategies on HNSW index for policy enforcement. The superscript L denotes impact on latency and superscript R denotes impact on Recall.}
\label{tab:policy_enforcement_comparison}

\footnotesize
\setlength{\tabcolsep}{3pt}

\begin{tabular}{p{3cm} p{2.5cm} p{2.5cm} p{2.5cm} p{1.3cm} p{2.5cm}}
\toprule
\textbf{Property} & \textbf{Pre-Filtering} & \textbf{Post-Filtering (PF)} & \textbf{Iterative PF} & \textbf{Parallel PF} & \textbf{Hybrid Filtering~\cite{acorn}} \\
\midrule

Policy Correctness & \high & \high & \high & \high & \high \\

Recall & \high & \low & \partialv & \partialv & \partialv \\

Query Latency & \high & \low & \partialv & \low & \partialv \\

Policy Selectivity 
& \high$^{L}$
& \high$^{R}$
& \partialv$^{R}$
& \partialv$^{R}$
& \high$^{L}$ \\

Policy Correlation 
& \nonev
& \high$^{R}$
& \high$^{Both}$
& \high$^{R}$
& \high$^{R}$ \\

Dynamic Policy 
& \low
& \low
& \partialv
& \high
& \partialv \\

Implementation Complexity 
& \low
& \low
& \low
& \partialv
& \high \\

\bottomrule
\end{tabular}

\end{table*}

\ifextended{
\noindent \textbf{Performance estimators.} 
We introduce analytical models for estimating recall and cost of a candidate strategy.
These models can be used by a query optimizer such that it selects the query execution strategy that minimizes latency while meeting the required recall. In this paper, we focus on pre-filtering and post-filtering strategies. 
\subsubsection*{Recall Model}
The maximum achievable recall by a strategy is given by number the vectors that satisfy the policies for a user $\PolicySet{S}$, \ie the policy selectivity: 
$Recall(s, \mathbb{Q}, P) = \Database \cdot \selectivity(P)$, where $s$ is pre-filtering strategy and $\selectivity(P)$ is the selectivity of policies.
In \emph{post-filtering}, 
recall depends on the underlying ANN index (e.g., HNSW), particularly graph connectivity $M$, entry-point search parameter $efs$, expansion factor $\beta$. These variables are provided as inputs to a linear regression model, which produces coefficient constants $\rho$ for each of the variable which are then used to compute recall.

\begin{equation*}
\text{recall}(s,\mathbb{Q},P) = \mathbf{x} \cdot \boldsymbol{\rho},
\quad \text{where } \mathbf{x} = (\beta, M, efs, K).
\end{equation*}

\subsubsection*{Cost model}

\showif{
The simplified regression model ignores the physical I/O metrics and approximates execution time as:

\begin{equation}
T_{scalar} = a + b_{SN} \times (S \times N) +  idx \times \text{index\_used}
\end{equation}

While this simplified model exhibits a reduced accuracy (\textbf{R\textsuperscript{2}} $\approx$ 0.27), it remains lightweight and suitable for use within the query optimizer, where only logical estimates are available prior to execution.
Following the scalar pre-filtering stage, the filtered records obtained from the scalar index scan are used as input for the vector similarity search stage. This second stage involves computing the distance between the query embedding and each of the filtered vectors to identify the top-$k$ nearest neighbors. 
To estimate the cost of this stage, we utilize the number of \textit{filtered rows} reported in the \texttt{EXPLAIN (ANALYZE, FORMAT JSON)} output of PostgreSQL. This value directly corresponds to the number of candidate vectors that participate in the distance computation process which is $\selectivity(\PolicySet{S})$ 
This count serves as the key driver of computational cost in the vector similarity search phase. The total cost of the vector stage is therefore modeled as a function of the number of distance calculations 
and the computational complexity of each operation. We formulate the execution time for this stage as:
\begin{equation}
T_{\text{distance-computations}} = a_v + b_v \times \rhovar(\PolicySet{S})
\end{equation}
The coefficients $a_v$ and $b_v$ are obtained empirically by executing queries over different selectivity conditions and measuring the observed vector search time. The resulting regression achieved a high goodness of fit (\textbf{R\textsuperscript{2}} $> 0.9$), confirming that the number of filtered vectors is the dominant factor influencing vector search cost.
Finally, the overall query cost is represented as a combination of the scalar pre-filtering and vector similarity stages.
}

Our first attempt at modeling this cost was based on a linear regression model.
While this model provided an interpretable analytical formulation for estimating vector query costs, its ability to capture complex non-linear relationships among parameters was limited.  
To address this limitation, we used a Random Forest Regression (RFR)–based cost model that empirically learns query runtime behavior from observed execution data. \cite{RandomForest}
The model is trained using the logarithm of the measured query execution time, $\log(\textit{time}_{ms})$, as the target variable. After training, the predicted costs are transformed back to the original scale by exponentiation. Formally, the model can be represented as:
\begin{equation*}
\hat{T}_{RF} = \exp \left( f_{RF}(\log K, \log(N \cdot s), \log ef_{Search}, \log \beta) \right)
\end{equation*}

where $f_{RF}$ denotes the non-linear regression constant that model  learned by the Random Forest, it is a learned non-linear function instead of a fixed equation.
It shows that the Random Forest learns the mapping in log space useful because query times can vary exponentially with parameters (especially efs and selectivity).

We trained the cost and recall models using execution time and recall data collected by running policies and vector queries on database to train two cost models: a scalar model for metadata filtering and a vector model for vector queries. The scalar model using linear regression achieved high accuracy (R² = 0.9), while the vector model performed poorly under linear regression but improved significantly with Random Forest Regression (R² = 0.8). When these models are fed to the optimizer, it generally selected the correct low-cost strategy with $<10$\% error, but mis-predicted pre-filtering in cases of high query-policy correlation. While these cost models are helpful, modeling additional factors such as high query-policy correlation is the next step to further improve accuracy. 

\showif{
\subsubsection*{Relevance Models}

Assuming relevance estimation across metadata filtering strategies is complex, we adopt a simplified ordering where Parallel Post-Filtering (PPF) achieves the highest relevance, followed by Iterative Post-Filtering (IPF), and then Pre-Filtering (PreF):
\[
\mathrm{Rel(k)}(\mathrm{PPF}) > \mathrm{Rel(k)}(\mathrm{IPF}) > \mathrm{Rel(k)}(\mathrm{PreF}).
\]
This reflects that post-filtering methods operate on larger candidate sets before applying policy constraints, helping preserve more relevant vectors, whereas pre-filtering restricts candidates early and may discard relevant results.
}

\fi

Table~\ref{tab:policy_enforcement_comparison} compares four policy enforcement strategies discussed in Section~\ref{subsec:policy-aware-vector-search-strategies}: pre-filtering, post-filtering, iterative filtering, and parallel filtering. The comparison considers key characteristics, including recall, correctness, policy selectivity, policy correlation, dynamic policy handling, implementation complexity and latency. Among these, recall, correctness, policy selectivity, and latency are derived from experimental results, against the ground-truth  top-$k$ results. Policy correlation, dynamic policy handling, and implementation complexity are characterized based on the expected behavior of each enforcement strategy. 
Overall, this comparison highlights the inherent trade-offs between each of the strategy when enforcing fine-grained access control in vector databases~\cite{HybridQuerying}. 
In Table~\ref{tab:policy_enforcement_comparison}, \textit{high}, \textit{partial}, and \textit{low} represent relative performance levels. For recall and policy correctness, they correspond to $>$90\%, 25–65\%, and $<$20\% of the ground truth, respectively. For latency, selectivity, and correlation, these labels indicate high, moderate, and low cost or the alignment relative to the highest observed values on ground truth. For dynamic policy support and implementation complexity, they reflect the degree of adaptability and system overhead, respectively.

Among these enforcement strategies, the system must select the strategy that guarantees policy-compliant results while achieving minimal possible query latency and maximal recall over the permitted data objects. Let $S$ denote the set of candidate enforcement strategies and let $\mathrm{Cost}(s, \mathbb{Q}, P)$ denote the execution cost of strategy $s \in S$ for a query $\mathbb{Q}$ and its associated FGAC policies $P$. We formalize this as an optimization problem: 

    \[
    s^{*}  =  \min_{s \in \mathcal{S}} \ \mathrm{Cost}(s, \mathbb{Q},P) \quad \text{subj. } \quad
        \begin{cases}
        \forall v \in R_{s}, \exists P_j \in P :\ P_j(v) = 1, \\[6pt]
        \mathrm{Recall}(R_s, \mathbb{Q}, V_P) \ge \tau.
        \end{cases}
    \]

Only strategies capable of achieving the user-specified recall threshold are considered feasible. 
This optimization framework allows the system to dynamically adapt its choice of metadata strategy based on both query characteristics and user requirements, ensuring an efficient balance between retrieval accuracy and execution performance~\footnote{To evaluate the correctness of any chosen strategy, the correctness criteria, proposed by Wang et al.~\cite{Correctness} for relational access control systems, has to be extended for vector databases. The three criteria are soundness, security, and maximality. Due to the semantic nature of vector search, they require substantial re-interpretation when used with vectors.}. In this context, we emphasize that policy correctness is strictly enforced in all strategies, ensuring that only authorized results are returned. The user-specified recall threshold serves solely as a workload-level quality constraint for selecting among enforcement strategies; it does not affect correctness. Correctness is enforced independently by ensuring that only authorized results are returned. Since vector search is inherently approximate, achieving full recall can be expensive. Therefore, recall is treated as a tunable parameter for balancing efficiency and result quality, and is used to identify feasible strategies.

\subsection{Strategies for Policy-aware Vector Search}
\label{subsec:policy-aware-vector-search-strategies}

In this section, we explain additional strategies for Policy-aware Vector Search beyond pre-filtering and post-filtering.

\noindent \textbf{Iterative Post Filtering (PF) }addresses the recall limitations of standard post-filtering by  expanding the search space. Instead of performing a single ANN search, this strategy performs iterative search: it repeatedly retrieves candidate vectors and applies scalar filters until enough authorized results are obtained. If the filtered results are insufficient to achieve top-$K$ result set, additional candidates are explored in subsequent iterations. This iterative expansion continues until the top-$K$ requirement is satisfied or a allocated memory limit is reached. In HNSW, this is controlled by increasing the candidate exploration parameter $hnsw.iterativescan$, enabling a larger portion of the graph to be traversed~\cite{surveyFANN}. 

\noindent \textbf{Parallel Post Filtering.}
We propose a parallel post-filtering approach to improve recall under selective policy constraints. Instead of expanding a single ANN search iteratively, this strategy  runs multiple ANN searches concurrently to retrieve diverse candidate subsets. This improves the chance of finding policy-compliant vectors, especially in low-correlation scenarios, where vectors similar to the query are unlikely to satisfy the policy filter.

In the sample dataset, we augmented each document with concept categories extracted from its title and abstract. These concepts provide an additional semantic layer for relating queries to policy constraints. We then generate concept-based query variants that remain close to the original query while guiding the search toward different regions of the vector space. This is useful in low-correlation scenarios, where vectors similar to the query are unlikely to satisfy the policy filter.

Parallel post-filtering executes the original query and its concept-based variants concurrently over the HNSW index. Each candidate set is filtered using the policy predicates, and the valid results are aggregated and de-duplicated. This allows the search to explore multiple semantic regions and improve the likelihood of retrieving sufficient authorized top-$K$ results. If the number of valid results meets the required top-$K$, the algorithm terminates. Otherwise, additional query variants are generated and the process is repeated.

%% file: sections/experiments.tex
\section{Preliminary Experiments}
We performed experiments on \pgsql with the \pgvector extension  \cite{pgvector} installed from binaries.  
We used the \textbf{arXiv} dataset, which includes fields such as \text{author}, \text{title}, \text{categories}, \text{license}, \text{abstract}, and others~\cite{dataset}.
It has a total count of  2,771,104 records with 4.74Gb size.
We used title and abstract as vector columns for embeddings,  rest of columns as metadata columns and sentence-transformer model 'all-MiniLM-L6-v2' \cite{all-minilm-l6-v2} to generate embeddings 
We compared four methods: pre-filtering (PF), na\"ive post-filtering (NPF), iterative post-filtering (IPF), and parallel post-filtering (PPF). For PPF we issued three parallel and medium selective queries to the HNSW index, ensuring each query explored a different region of the graph.  
To perform these experiments, we designed three different policy templates, combining multiple metadata columns available in the dataset. 
The workload generator instantiates these templates into concrete policies and ensures that each policy produces a non-empty result set.
Experimental results are evaluated by measuring recall and latency across different levels of policy selectivity. Policy selectivity represents the fraction of records that satisfy a given policy filter. In our experiments, we control selectivity by specifying the number of rows in the dataset that are allowed to pass the policy filter.
\subsubsection*{Experiment 1: Recall}
\begin{figure}[t]
    \centering
    \begin{subfigure}[b]{0.23\textwidth}
        \centering
        \includegraphics[width=\linewidth]{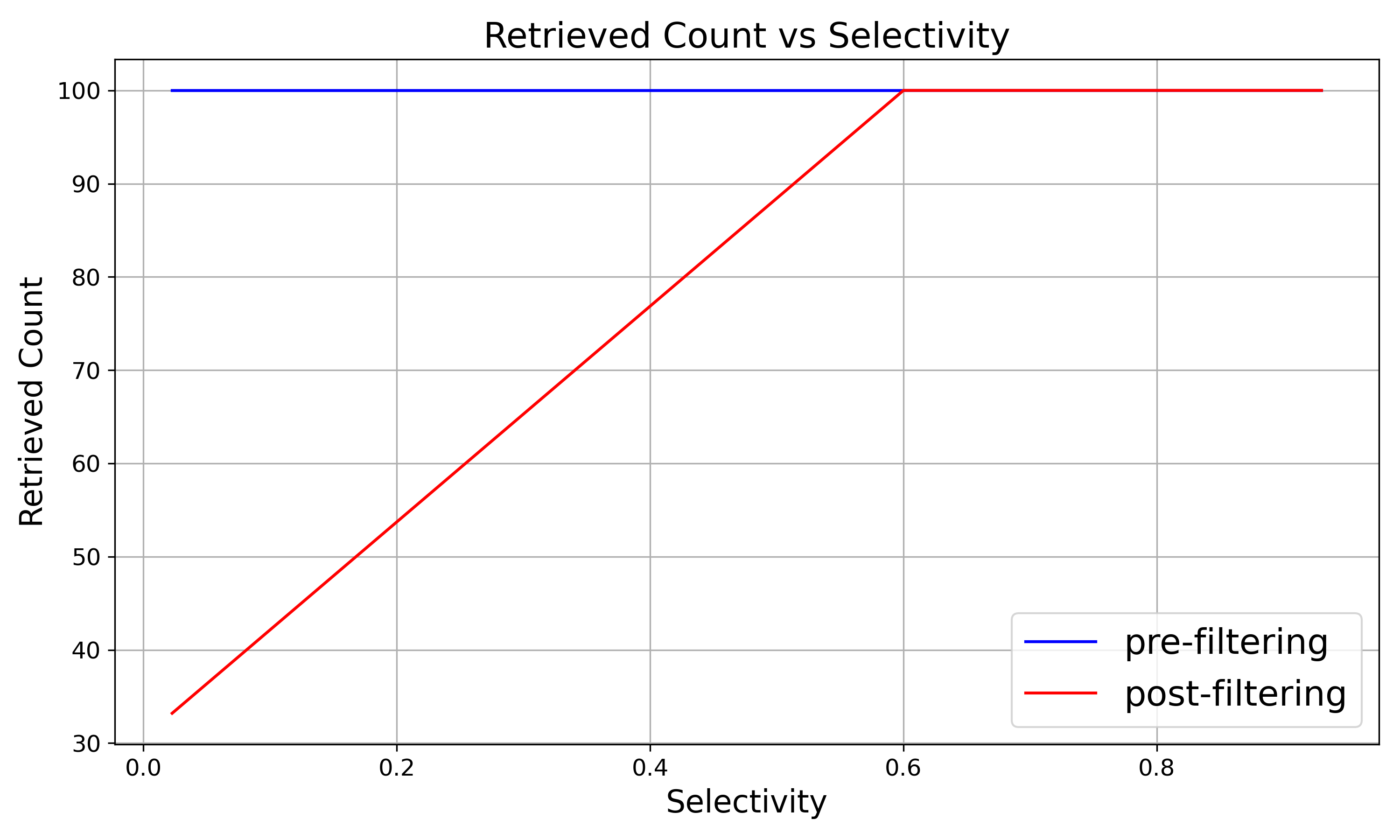}
        \caption{\footnotesize Default Recall: PF vs. NPF}
        \label{fig:Default_recall}
    \end{subfigure}
    \hfill
    \begin{subfigure}[b]{0.23\textwidth}
        \centering
        \includegraphics[width=\linewidth]{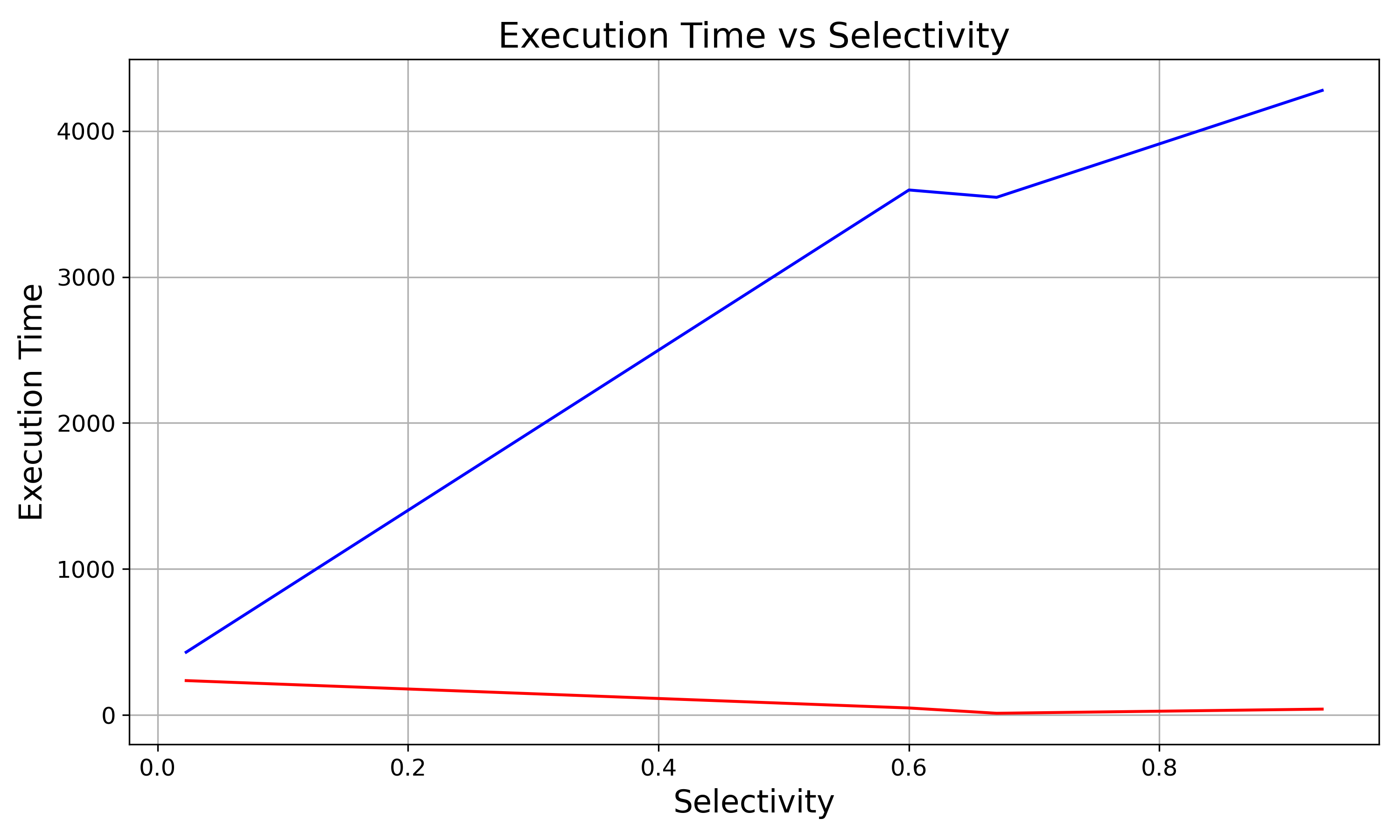}
        \caption{\footnotesize Default Latency: PF vs. NPF }
        \label{fig:Default_latency}
    \end{subfigure}
    \hfill
    \begin{subfigure}[b]{0.23\textwidth}
        \centering
        \includegraphics[width=\linewidth]{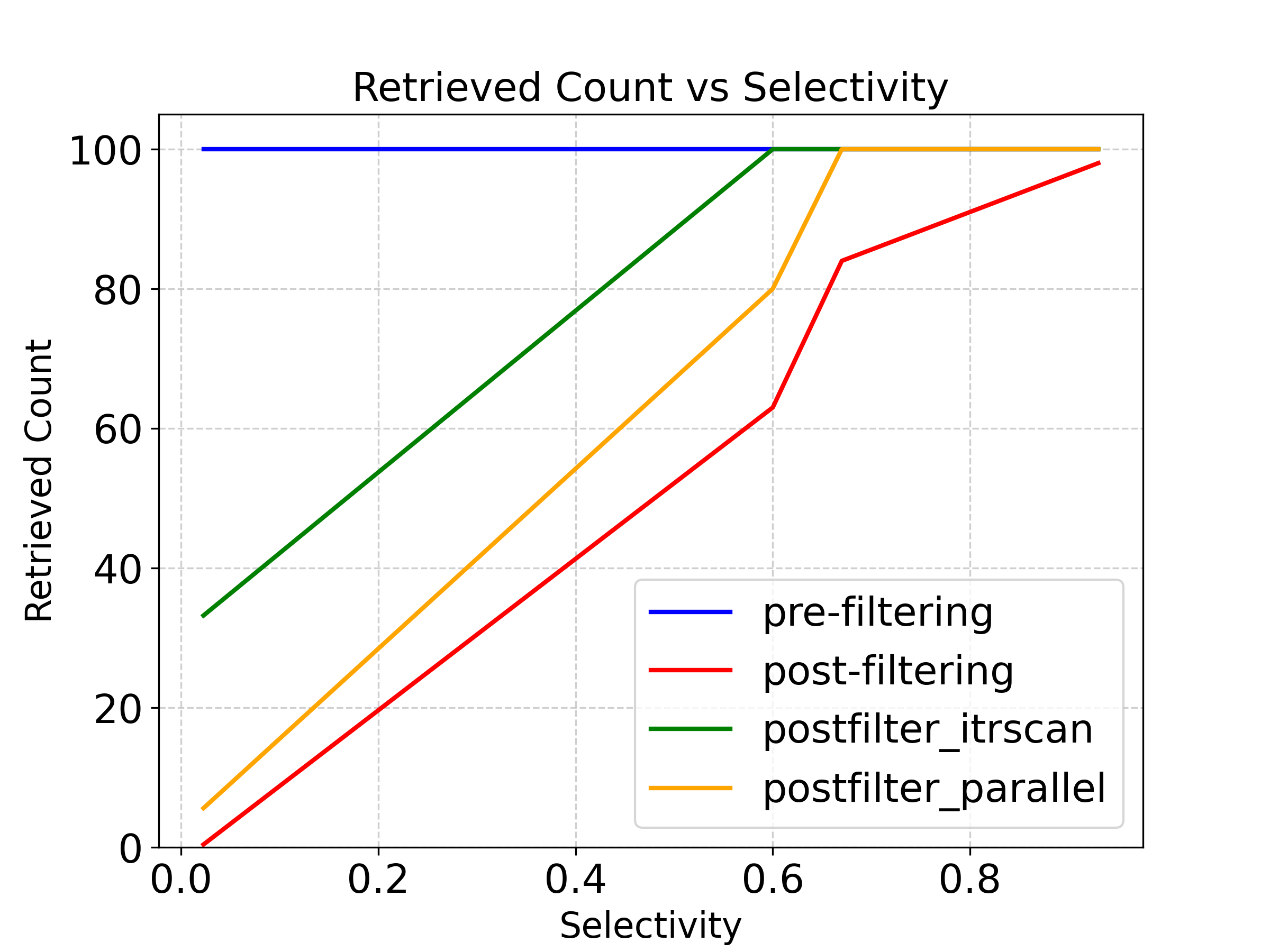}
        \caption{\footnotesize Recall for all 4 strategies }
        \label{fig:Parallel_recall}
    \end{subfigure}
    \hfill
    \begin{subfigure}[b]{0.23\textwidth}
        \centering
        \includegraphics[width=\linewidth]{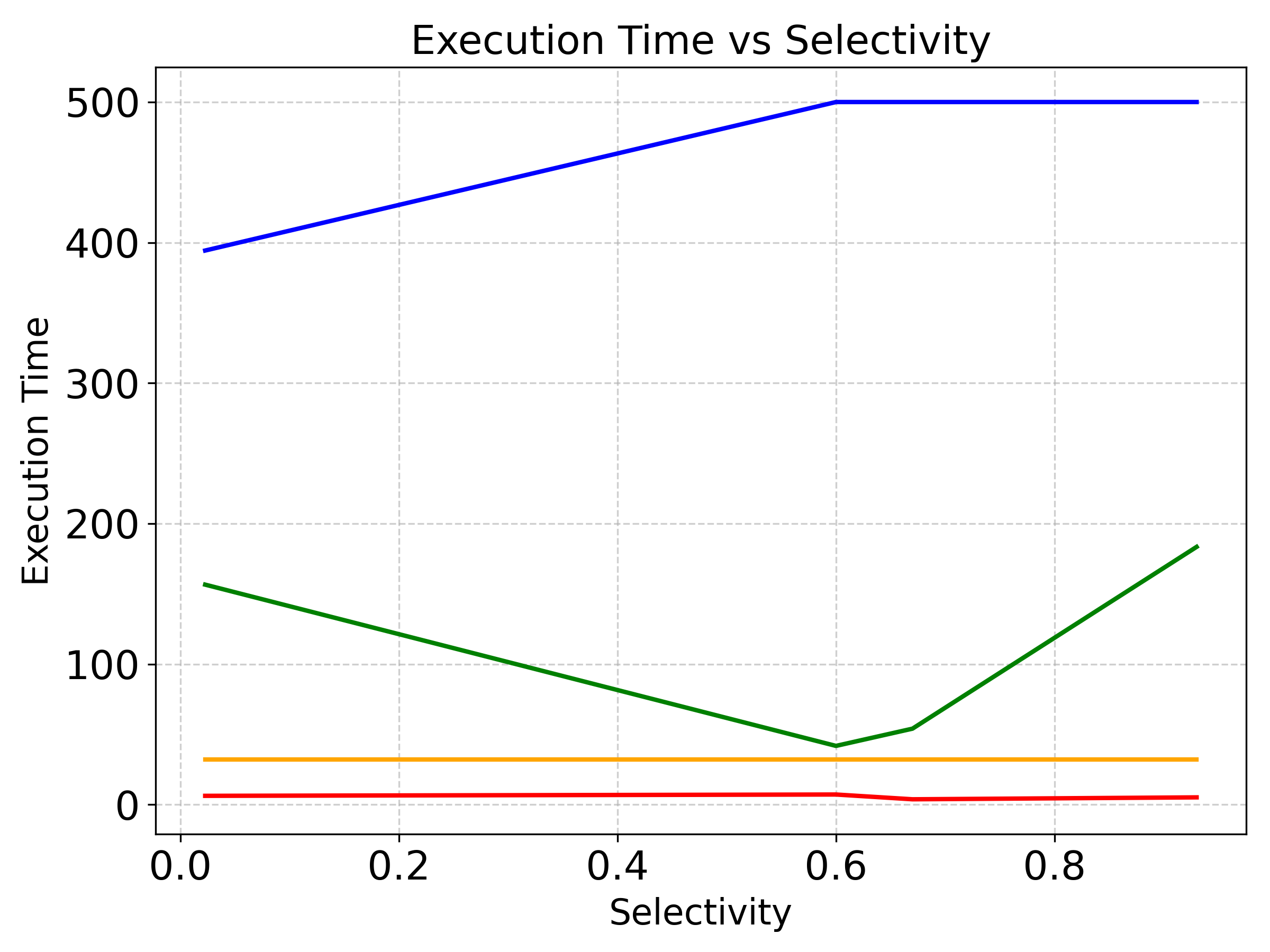}
        \caption{\footnotesize Latency for all 4 strategies}
        \label{fig:Parallel_latency}
    \end{subfigure}

    \caption{Recall and Latency versus Selectivity for different enforcement strategies.}
    \Description[Experimental results comparing different strategies]{Recall and Latency versus Selectivity for different enforcement strategies.}
\end{figure}

We evaluated the default \pgvector behavior for both PF and NPF across all policy templates. As shown in Fig.~\ref{fig:Default_recall}, pre-filtering consistently achieved highest recall regardless of selectivity, while post-filtering generally produced lower recall, except in cases where the policy predicate was highly correlated with the user query.
In the next experiment, we evaluated recall-versus-selectivity for all 4 strategies as shown in Fig.~\ref{fig:Parallel_recall}. PPF achieved competitive recall for medium-selective workloads, outperforming na\"ive post-filtering. Notably, this method achieved these recall improvements while reducing execution time compared to the iterative scan based post-filtering approach as shown next.

\subsubsection*{Experiment 2: Latency} 
This experiment compares the execution time of different enforcement strategies across different policy selectivity levels. The na\"ive post-filtering strategy exhibits the lowest execution time overall but it retrieves very few policy-compliant vectors, resulting in almost zero recall. 
Pre-filtering incurs higher execution time under low-selectivity policies because a large candidate set is passed to the vector search, offering little reduction in the distance computations performed. Conversely, under high-selectivity policies, pre-filtering becomes more efficient as the reduced candidate set meaningfully constrains the search space.
The same experiment when iterative scan is enabled (Fig.~\ref{fig:Parallel_latency}), post-filtering execution time rises sharply. Each additional scan round retrieves another batch of candidates, increasing cost substantially for low and medium selective policies.  In the final comparison (Fig.~\ref{fig:Parallel_latency}), parallel post-filtering reduces the number of search rounds required by iterative post-filtering. Instead it explored different regions of the HNSW index. This strategy delivered lower latency than iterative scan, and better recall compared to naive post-filtering for medium selectivity ranges.

%% file: sections/discussion.tex
\section{Discussion}

In this section, we briefly outline the differences between FGAC \& FANN and some of the possible future research directions to make \textit{Policy-aware vector research} a reality.

\noindent \textbf{FGAC vs FANN.}
Fine-grained access control (FGAC) in vector databases fundamentally differs from traditional filtered approximate nearest neighbor (FANN) search. In FANN systems, filters are typically broad, low-selectivity metadata predicates attached directly to the query~\cite{surveyFANN}. In contrast, FGAC enforces user-specific authorization policies that may contain highly selective and complex predicate combinations (e.g., CNF/DNF policies). Unlike FANN, where the primary goals are retrieval latency and high recall, correctness of enforcement becomes the primary requirement in FGAC, since unauthorized vector retrieval can lead to sensitive data leakage. While FGAC mechanisms can support traditional filtered ANN workloads, it is unclear whether solutions designed only for FANN are sufficient for access-controlled retrieval. 

\noindent \textbf{Cost Models for Enforcement Strategies.}
Our preliminary experiments on cost models to estimate query latency and recall and to guide the choice between pre-filtering and post-filtering strategies, show that highly selective predicates tend to favor pre-filtering, whereas low-selectivity scenarios benefit from post-filtering. However, many real workloads fall between these extremes.  
Hybrid filtering strategies can help in the intermediate stage where predicate selectivity is neither extremely high nor extremely low~\cite{acorn, UNG}. A unified cost model that dynamically differentiates among all possible strategies can significantly improve plan selection, leading to more reliable latency and recall guarantees for fine-grained access control workloads. Developing such a unified cost and recall models for different strategies is an open research problem. 

\ifextended
\noindent \textbf{Policy Storage and Maintenance.}
Policies formulated over metadata are stored in a relational table along with a unique policy ID, and are retrieved based on subject constraints. 
When policies are getting updated more frequently—such as adding, removing, or modifying policy constraints the policy-subject constraints table needs to be updated. In this case, if the modified policies apply to a different subset of the data, varying the selectivity or if introduce a higher variance in other model parameters, additional training data may be required. This introduces additional overhead for re-evaluating policies, updating models, or reapplying policy–subject constraint checks to ensure correctness. Moreover, in vector databases, where approximate nearest neighbor (ANN) search is commonly used, policy changes can affect the candidate search space and potentially impact both recall and query latency.
\fi 

\noindent \textbf{Enforcing Vector-based Object Conditions.}
In this paper, we only discussed enforcement of metadata-based object conditions. Enforcing vector-based object conditions requires a different approach for supporting two independent vector queries. The correct yet costly approach could involve a dual-index strategy, employing an exact-search index (e.g., IVF-Flat) for policy vectors alongside an ANN index for query vectors, and returning the intersection of the resulting sets. An offline policy subgraph construction approach where policy-compliant vectors are pre-clustered into navigable subgraphs, similiar to ACORN~\cite{acorn}, can structurally confine query-time search to the authorized region of the index without post-hoc filtering. A joint embedding approach that fuses policy and query embeddings into a unified representation prior to retrieval, collapsing enforcement and semantic search into a single vector operation.

%% file: sections/conclusions.tex
\section{Conclusions \&  Future Work}
Our goal is to treat Fine Grained Access Control as a first-class concern in vector databases, jointly modeling metadata and vector-level policies and introducing a cost-based framework that dynamically selects efficient enforcement strategies while meeting latency and recall requirements. 
To achieve this vision, access control and approximate nearest neighbor search must be co-designed from the ground up, rather than treating enforcement as a purely post-hoc layer to existing strategies and indexes. 
Evaluating the proposed approaches across multiple database systems and a diverse set of datasets remains as future work, to validate their robustness under varying data distributions and different system characteristics.

\section{Acknowledgment} 
This work is partially supported by the NSF Award \#2451803.

%% file: references.bib
@article{HybridQuerying,
author = {Zhu, Jiaxu and Yuan, Jiayu and Yang, Kaiwen and Chen, Xiaobao and Yu, Shihuan and Lv, Hongchang and Li, Yan and Zheng, Bolong},
title = {An Experimental Evaluation of Hybrid Querying on Vectors},
year = {2025},
issue_date = {October 2025},
publisher = {VLDB Endowment},
volume = {19},
number = {2},
issn = {2150-8097},
url = {https://doi.org/10.14778/3773749.3773757},
doi = {10.14778/3773749.3773757},
journal = {Proc. VLDB Endow.},
month = oct,
pages = {183–195},
numpages = {13}
}

@misc{surveyFANN,
      title={Survey of Filtered Approximate Nearest Neighbor Search over the Vector-Scalar Hybrid Data}, 
      author={Yanjun Lin and Kai Zhang and Zhenying He and Yinan Jing and X. Sean Wang},
      year={2025},
      eprint={2505.06501},
      archivePrefix={arXiv},
      primaryClass={cs.DB},
      url={https://arxiv.org/abs/2505.06501}, 
}

@article{acorn,
  title={Acorn: Performant and predicate-agnostic search over vector embeddings and structured data},
  author={Patel, Liana and Kraft, Peter and Guestrin, Carlos and Zaharia, Matei},
  journal={Proceedings of the ACM on Management of Data},
  volume={2},
  number={3},
  pages={1--27},
  year={2024},
  publisher={ACM New York, NY, USA}
}

@article{hnsw,
  title={Efficient and robust approximate nearest neighbor search using hierarchical navigable small world graphs},
  author={Malkov, Yu A and Yashunin, Dmitry A},
  journal={IEEE transactions on pattern analysis and machine intelligence},
  volume={42},
  number={4},
  pages={824--836},
  year={2018},
  publisher={IEEE}
}

@article{sieve,
  author       = {Primal Pappachan and
                  Roberto Yus and
                  Sharad Mehrotra and
                  Johann{-}Christoph Freytag},
  title        = {Sieve: {A} Middleware Approach to Scalable Access Control for Database
                  Management Systems},
  journal      = {Proc. {VLDB} Endow.},
  volume       = {13},
  number       = {11},
  pages        = {2424--2437},
  year         = {2020},
  url          = {http://www.vldb.org/pvldb/vol13/p2424-pappachan.pdf},
  bibsource    = {dblp computer science bibliography, https://dblp.org}
}

@article{ABAC,
	author = {Aserto},
	title = {ABAC-RBAC},
	journal = {\url{https://www.aserto.com/blog/rbac-vs-abac-authorization-models }},
	year = {},
	note = {},
}

@article{Filtered-DiskANN,
author = {Gollapudi, Siddharth and Karia, Neel and Sivashankar, Varun and Krishnaswamy, Ravishankar and Begwani, Nikit and Raz, Swapnil and Lin, Yiyong and Zhang, Yin and Mahapatro, Neelam and Srinivasan, Premkumar and Singh, Amit and Simhadri, Harsha Vardhan},
title = {Filtered-DiskANN: Graph Algorithms for Approximate Nearest Neighbor Search with Filters},
year = {2023},
isbn = {9781450394161},
publisher = {Association for Computing Machinery},
address = {New York, NY, USA},
url = {https://doi.org/10.1145/3543507.3583552},
journal = {10.1145/3543507.3583552},
booktitle = {Proceedings of the ACM Web Conference 2023},
pages = {3406–3416},
numpages = {11},
location = {Austin, TX, USA},
series = {WWW '23}
}

@dataset {dataset,
author = {},
title = {arXiv Dataset},
license = {CC0: Public Domain},
size = {4.56 Gb},
url = {https://www.kaggle.com/datasets/Cornell-University/arxiv}
}

@article{honeybee,
author = {Zhong, Hongbin and Lentz, Matthew and Narodytska, Nina and Szekeres, Adriana and Rong, Kexin},
title = {HONEYBEE: Efficient Role-based Access Control for Vector Databases via Dynamic Partitioning},
year = {2026},
issue_date = {February 2026},
publisher = {Association for Computing Machinery},
address = {New York, NY, USA},
volume = {4},
number = {1},
url = {https://doi.org/10.1145/3786625},
doi = {10.1145/3786625},
journal = {Proc. ACM Manag. Data},
month = apr,
articleno = {11},
numpages = {26}
}

@article{UNG,
author = {Cai, Yuzheng and Shi, Jiayang and Chen, Yizhuo and Zheng, Weiguo},
title = {Navigating Labels and Vectors: A Unified Approach to Filtered Approximate Nearest Neighbor Search},
year = {2024},
issue_date = {December 2024},
publisher = {Association for Computing Machinery},
address = {New York, NY, USA},
volume = {2},
number = {6},
url = {https://doi.org/10.1145/3698822},
doi = {10.1145/3698822},
journal = {Proc. ACM Manag. Data},
month = dec,
articleno = {246},
numpages = {27}
}

@misc{RandomForest,
  author       = {Utsav Pathak and Amit Mankodi},
  title        = {Redefining Cost Estimation in Database Systems: The Role of Execution Plan Features and Machine Learning},
  howpublished = {arXiv preprint arXiv:2510.05612v1},
  year         = {2025},
  url          = {https://arxiv.org/abs/2510.05612v1}
}

@inproceedings{Correctness,
author = {Wang, Qihua and Yu, Ting and Li, Ninghui and Lobo, Jorge and Bertino, Elisa and Irwin, Keith and Byun, Ji-Won},
title = {On the correctness criteria of fine-grained access control in relational databases},
year = {2007},
isbn = {9781595936493},
publisher = {VLDB Endowment},
pages = {555–566},
numpages = {12},
location = {Vienna, Austria},
series = {VLDB '07}
}

@INPROCEEDINGS{PBAC,
  author={Yang, Naikuo and Barringer, Howard and Zhang, Ning},
  booktitle={Third International Symposium on Information Assurance and Security}, 
  title={A Purpose-Based Access Control Model}, 
  year={2007},
  volume={},
  number={},
  pages={143-148},
  doi={10.1109/IAS.2007.29}}

@misc{pgvector,
  author = {Andrew Kane},
  title = {pgvector: Open-source vector similarity search for Postgres},
  year = {2024},
  publisher = {GitHub},
  journal = {GitHub repository},
  howpublished = {\url{https://github.com/pgvector/pgvector}},
  commit = {<Add the specific commit hash used, e.g., 8e1b2c>}
}

@software{all-minilm-l6-v2,
  author = {Reimers, Nils},
  title = {Sentence-Transformers: all-MiniLM-L6-v2},
  url = {https://huggingface.co/sentence-transformers/all-MiniLM-L6-v2},
  year = {2021}
}

@article{tattletale,
 author = {Primal Pappachan and
Shufan Zhang and
Xi He and
Sharad Mehrotra},
 bibsource = {dblp computer science bibliography, https://dblp.org},
 journal = {Proc. VLDB Endow.},
 number = {11},
 pages = {2437--2449},
 title = {Don't Be a Tattle-Tale: Preventing Leakages through Data Dependencies
on Access Control Protected Data},
 url = {https://www.vldb.org/pvldb/vol15/p2437-pappachan.pdf},
 volume = {15},
 year = {2022}
}
